\documentclass[final,5p,times,twocolumn,sort&compress]{elsarticle}
\usepackage{amssymb}
\usepackage{amsmath}
\usepackage{graphicx}
\usepackage{lscape}
\usepackage{booktabs,longtable}
\usepackage{mathptmx, courier, pifont}
\usepackage[scaled=0.92]{helvet}
\usepackage[T1]{fontenc}
\usepackage{textcomp}
\usepackage{rotating}
\usepackage{verbatim}
\usepackage{picinpar}
\usepackage{wrapfig}
\usepackage{float}
\usepackage{multirow}
\usepackage{graphicx}
\usepackage[dvipsnames]{xcolor}
\usepackage[colorlinks=true,urlcolor=Blue,linkcolor=Blue]{hyperref}
\usepackage[all]{hypcap}
\usepackage{color}

\usepackage{datetime}
\journal{Physics Letters B}

\begin{document}
\begin{frontmatter}

\title{Mass Measurements of Neutron-Deficient Y, Zr, and Nb Isotopes \\ and Their Impact on $rp$ and \emph{$\nu$p} Nucleosynthesis Processes}
\date{today}
\author[IMP,GSI]{Y.~M.~Xing}%
\author[IMP]{K.~A.~Li}%
\author[IMP,EMMI]{Y.~H.~Zhang}%
\author[IMP]{X.~H.~Zhou\fnref{1}}\fntext[1]{Corresponding author. Email address: zxh@impcas.ac.cn}
\author[IMP]{M.~Wang\fnref{2}}\fntext[2]{Corresponding author. Email address: wangm@impcas.ac.cn}
\author[IMP,GSI]{Yu.~A.~Litvinov\fnref{3}}\fntext[3]{Corresponding author. Email address: y.litvinov@gsi.de}
\author[MPIK]{K.~Blaum}%
\author[Sophia,iTHES]{S.~Wanajo}%
\author[IMP,Kubono]{S.~Kubono\fnref{4}}\fntext[4]{Corresponding author. Email address: kubono@riken.jp}
\author[GSI,TUD]{G.~Mart{\'i}nez-Pinedo}%
\author[TUD,GSI]{A.~Sieverding}%
\author[IMP,GSI]{R.~J.~Chen}%
\author[IMP]{P.~Shuai}%
\author[IMP,UNI]{C.~Y.~Fu}%
\author[IMP]{X.~L.~Yan}
\author[Orsay]{W.~J.~Huang}%
\author[IMP]{X.~Xu}%
\author[IMP]{X.~D.~Tang}%
\author[IMP]{H.~S.~Xu}%
\author[IMP]{T.~Bao}%
\author[IMP]{X.~C.~Chen}%
\author[IMP]{B.~S.~Gao}%
\author[IMP,He]{J.~J.~He}%
\author[IMP]{Y.~H.~Lam}%
\author[IMP,UNI]{H.~F.~Li}%
\author[IMP,UNI]{J.~H.~Liu}%
\author[IMP]{X.~W.~Ma}%
\author[IMP]{R.~S. Mao}%
\author[IMP,UNI]{M.~Si}%
\author[IMP,UNI]{M.~Z.~Sun}%
\author[IMP]{X.~L.~Tu}%
\author[IMP,UNI]{Q.~Wang}%
\author[IMP]{J.~C.~Yang}%
\author[IMP]{Y.~J.~Yuan}%
\author[IMP,Zeng]{Q.~Zeng}%
\author[IMP,UNI]{P.~Zhang}%
\author[IMP,UNI]{X.~Zhou}%
\author[IMP]{W.~L.~Zhan}%
\author[GSI]{S.~Litvinov}%
\author[Orsay]{G.~Audi}%
\author[RIKEN]{T.~Uesaka}%
\author[RIKEN]{Y.~Yamaguchi}%
\author[UniS]{T.~Yamaguchi}%
\author[UniT]{A.~Ozawa}%
\author[Carla]{C.~Fr{\"o}hlich}%
\author[Tommi]{T.~Rauscher}%
\author[Friedel]{F.-K.~Thielemann}%
\author[Baohua]{B.~H.~Sun}%
\author[Sun]{Y.~Sun}%
\author[Furong]{A.~C.~Dai}%
\author[Furong]{F.~R.~Xu}%

\address[IMP]{CAS Key Laboratory of High Precision Nuclear Spectroscopy and Center for Nuclear Matter Science, Institute of Modern Physics, Chinese Academy of Sciences, Lanzhou 730000, People's Republic of China}%
\address[GSI]{GSI Helmholtzzentrum f{\"u}r Schwerionenforschung, Planckstra{\ss}e 1, Darmstadt, 64291 Germany}%
\address[EMMI]{ExtreMe Matter Institute EMMI, GSI Helmholtzzentrum f{\"u}r Schwerionenforschung, Planckstra{\ss}e 1, 64291 Darmstadt, Germany}%
\address[MPIK]{Max-Planck-Institut f\"{u}r Kernphysik, Saupfercheckweg 1, 69117 Heidelberg, Germany}%
\address[Sophia]{Department of Engineering and Applied Sciences, Sophia University, Chiyoda-ku, Tokyo 102-8554, Japan}%
\address[iTHES]{iTHES Research Group, RIKEN, Wako, Saitama 351-0198, Japan}%
\address[Kubono]{Center for Nuclear Study, University of Tokyo, Saitama 351-0198, Japan}%
\address[TUD]{Institut f{\"u}r Kernphysik (Theoriezentrum), Technische Universit{\"a}t Darmstadt, Schlossgartenstra{\ss}e 2, 64289 Darmstadt, Germany}%
\address[UNI]{University of Chinese Academy of Sciences, Beijing 100049, People's Republic of China}%
\address[Orsay]{CSNSM, Univ Paris-Sud, CNRS/IN2P3, Universit\'{e} Paris-Saclay, 91405 Orsay, France}%
\address[He]{Key Laboratory of Optical Astronomy, National Astronomical Observatories, Chinese Academy of Sciences, Beijing 100012, People's Republic of China}%
\address[Zeng]{Research Center for Hadron Physics, National Laboratory of Heavy Ion Accelerator Facility in Lanzhou and University of Science and Technology of China, Hefei 230026, People's Republic of China}%
\address[RIKEN]{RIKEN Nishina Center, RIKEN, Saitama 351-0198, Japan}%
\address[UniS]{Department of Physics, Saitama University, Saitama 338-8570, Japan}%
\address[UniT]{Insititute of Physics, University of Tsukuba, Ibaraki 305-8571, Japan}%
\address[Carla]{Department of Physics, North Carolina State University, Raleigh, NC 27695-8202, USA}%
\address[Tommi]{Centre for Astrophysical Research, University of Hertfordshire, Hatfield AL10 9AB, UK}%
\address[Friedel]{Department of Physics, University of Basel, Klingelbergstra{\ss}e 82, CH-4056 Basel, Switzerland}%
\address[Baohua]{School of Physics and Nuclear Energy Engineering, Beihang University, Beijing 100191, People's Republic of China}%
\address[Sun]{Department of Physics and Astronomy, Shanghai Jiao Tong University, Shanghai 200240, People's Republic of China}%
\address[Furong]{State Key Laboratory of Nuclear Physics and Technology, School of Physics, Peking University, Beijing 100871, People's Republic of China}%

\begin{abstract}
Using isochronous mass spectrometry at the experimental storage ring CSRe in Lanzhou,
the masses of $^{82}$Zr and $^{84}$Nb were measured for the first time with an uncertainty of $\sim 10$ keV,
and the masses of $^{79}$Y, $^{81}$Zr, and $^{83}$Nb were re-determined with a higher precision.
The latter are significantly less bound than their literature values.
Our new and accurate masses remove the irregularities of the mass surface in this region of the nuclear chart. Our results do not support the predicted island of pronounced low $\alpha$ separation energies for neutron-deficient Mo and Tc isotopes,
making the formation of Zr-Nb cycle in the $rp$-process unlikely.
The new proton separation energy of $^{83}$Nb was determined to be 490(400)~keV smaller than that in the Atomic Mass Evaluation 2012. This  partly removes the overproduction of the $p$-nucleus $^{84}$Sr relative to the neutron-deficient molybdenum isotopes in the previous \emph{$\nu$p}-process simulations.

\end{abstract}


\end{frontmatter}%

\section{Introduction}

Stellar nucleosynthesis, especially of heavy nuclei, is of great interest in nuclear astrophysics~\cite{Rauscher2014,Kajino2016}.
There are two well-known scenarios for producing almost all heavy chemical elements beyond Fe called $s$-process~\cite{Kajino2016,Kappeler2011} and $r$-process~\cite{Kajino2016,Arnould2007}.
Besides, there is a third class of stable nuclei categorized as $p$-nuclei~\cite{Kajino2016,Arnould2003} which amount to less than one percent of the elemental abundances above Z$\geq$34.
Most of the heavy $p$-nuclei can be explained by the photo-dissociation ($\gamma$-process) under high-temperature environments in supernova explosions. However, the "light $p$-nuclei" in the medium mass region, whose abundance ratios are anomalously large, cannot be understood in the framework of standard nucleosynthesis~\cite{Rauscher2002,Arnould2003}.
Natural consideration is a contribution from explosive burning through hydrogen-rich region, like the $rp$-~\cite{Wallace1981} or $\nu p$-~\cite{Frohlich2006,Wanajo2006,Pruet2006} processes. 
Recently, a possibility of a synthesis in the Type Ia supernovae was also reported~\cite{Trav2011}. 
Specifically, we concentrate here on the study of possible contributions of the $rp$-process and $\nu p$-process for
the light $p$-nuclei around $A\sim 90-100$, including  $^{92,94}$Mo and $^{84}$Sr~\cite{Woosley1990,Frohlich2006,Wanajo2006,Pruet2006}. Here, $^{84}$Sr is also considered to be anomalously abundant $p$-nuclide~\cite{Wanajo2011}, as the absolute abundance is comparable to those of $^{92,94}$Mo.

The astrophysical $rp$-~\cite{Wallace1981} and $\nu p$-~\cite{Frohlich2006,Wanajo2006,Pruet2006} processes
have been suggested~\cite{Schatz98,Schatz2001} to describe the production of light $p$-nuclei.
The former is related to type I X-ray bursts which occur on the surface of neutron stars accreting H- and He-rich matter from a companion star in a stellar binary system.
The bursts appear periodically in hours or days corresponding to the matter-accumulation time and last for tens to hundreds of seconds.
During this time neutron-deficient nuclei up to the Sn region~\cite{Schatz2001,Elomaa} can be synthesized via a sequence of proton captures and $\beta^+$ decays.
Although there is still a debate on the contribution of type I X-ray bursts to the galactic element abundances~\cite{Parikh2013}, such scenarios can not be totally excluded.
The $\nu p$-process is considered to occur in the inner ejecta of core-collapse supernovae which last for less than 10~s \cite{Frohlich2006,Wanajo2006,Pruet2006}.
Here, slow $\beta^+$ decays of the waiting point nuclei are replaced by fast $(n,p)$ reactions,
where neutrons are produced in reactions of electron anti-neutrinos in the neutrino winds with free protons in the ejecta.
The \emph{$\nu$p}-process can produce light $p$-nuclei up to $A\sim110$ including $^{92,94}$Mo, $^{96,98}$Ru, and $^{84}$Sr~\cite{Frohlich2006,Wanajo2006,Pruet2006}. All the $\nu p$-process simulations predict quite high production of $^{84}$Sr, which might be due to insufficiently known nuclear data and/or $\nu p$-process scenarios.

Although both, $rp$- and $\nu p$-, processes are sensitive to the physical conditions of the stellar environments \cite{Arcones1,Wanajo2011},
nuclear physics parameters, especially the atomic masses of nuclides
along the reaction paths, play a crucial role~\cite{Parikh2009,Schatz2016,Gabriel-Book,Wanajo2011}.
On the one hand, $rp$-process model calculations based on the finite range droplet mass model 1992 (FRDM$^{\prime }$92)~\cite{moller95} predicted the formation of a Zr-Nb cycle~\cite{Schatz98}.
The latter has been emphasized recently based on the previous experimental mass values because it would impose an upper temperature limit for the synthesis of elements beyond Nb~\cite{Haettner2011}.
On the other hand, the production of light $p$-nuclei in the $\nu p$-process relies on unknown or highly
uncertain
masses in the $A=79-84$ region.
In particular precise masses~\cite{Fallis2008,Fisker2009,Fallis2011,Weber2008} of nuclei along
the path
are important to explain the
observed solar abundances of $^{92}$Mo and $^{94}$Mo~\cite{Fallis2008,Weber2008}.
By taking the data from the 2003 Atomic Mass Evaluation (AME$^{\prime}03$)~\cite{Audi2003},
it has already been shown that
masses of $^{82}$Zr and $^{83}$Nb are crucial for
the production of $^{84}$Sr~\cite{Frohlich2006,Wanajo2011}.
Learning about the contribution of the $\nu p$-process to
$^{84}$Sr can be decisive in understanding the origin of the most
mysterious $p$-nuclei $^{92,94}$Mo~\cite{Wanajo2011}.


In this Letter, we report on precision mass measurements of five nuclei around A$\sim$79$-$84.
We address the region of low $\alpha$-separation energies predicted by FRDM$^{\prime}$92 in neutron-deficient Mo and Tc isotopes and
conclude on an impossible existence of the Zr-Nb cycle in the $rp$-process. Furthermore, we discuss
the overproduction of $^{84}$Sr in the \emph{$\nu$p}-process.
%

\section{Experiment}

The experiment was conducted at the HIRFL-CSR accelerator complex~\cite{Xiajiawen,zwl2010} in the Institute of Modern Physics in Lanzhou.
It was done in a similar way to our previous measurements described in Refs.~\cite{Tu2011, Shuai-MFC,Chen2017}.
Therefore only a brief description and specific details are given here.

A 400~MeV/u $^{112}$Sn$^{35+}$ primary beam of about $8\times 10^7$ particles per spill
was delivered by the heavy-ion synchrotron CSRm and focused upon a $\sim $10~mm $^9$Be target placed at the entrance of the fragment separator RIBLL2.
The reaction products from projectile fragmentation of $^{112}$Sn emerged from the target mainly as bare ions.
They were analyzed in flight~\cite{Geis92} by RIBLL2.
A cocktail beam including the ions of interest was then injected into the experimental storage ring CSRe.
The RIBLL2-CSRe were set to $B\rho=5.3347$~Tm corresponding to the maximum transmission for $^{101}$In.
The CSRe was tuned into the isochronous ion-optical mode with the transition point set to $\gamma_t=1.302$.
In this mode the revolution times of the ions depend in first order only on their mass-to-charge ratios \cite{ims1,ims2,ims3,ims4}.

A dedicated timing detector~\cite{Mei2010} was installed inside the CSRe aperture.
It was equipped with a 19 $\mu$g/cm$^2$ carbon foil of 40 mm in diameter.
Each time when an ion passed through the foil, secondary electrons were released from the foil surface.
The electrons were transmitted isochronously by perpendicularly arranged electric and magnetic fields to
a micro-channel plate (MCP) counter.
The signals from the MCP were guided without amplification directly to a fast digital oscilloscope.
The detection efficiency of the detector
varied from about 20\% to 80\% depending on the overall number of stored ions and their charge.
For each injection, a measurement time of $200$~$\mu$s, triggered by the CSRe injection kicker, was set corresponding to about $300$ revolutions of the ions in the CSRe.

Considered in the analysis were the ions which satisfied two requirements simultaneously: (1) at least 40 time signals were recorded for each ion and (2) the ion should circulate in the ring for more than 50 $\mu$s. The revolution time spectrum was obtained from all injections analogously to our previous analyses, details of which can be found in Refs.~\cite{Tu2011, Shuai-MFC,Chen2017}.


 \begin{figure*}[t]
 \centering
\includegraphics[width=15cm]{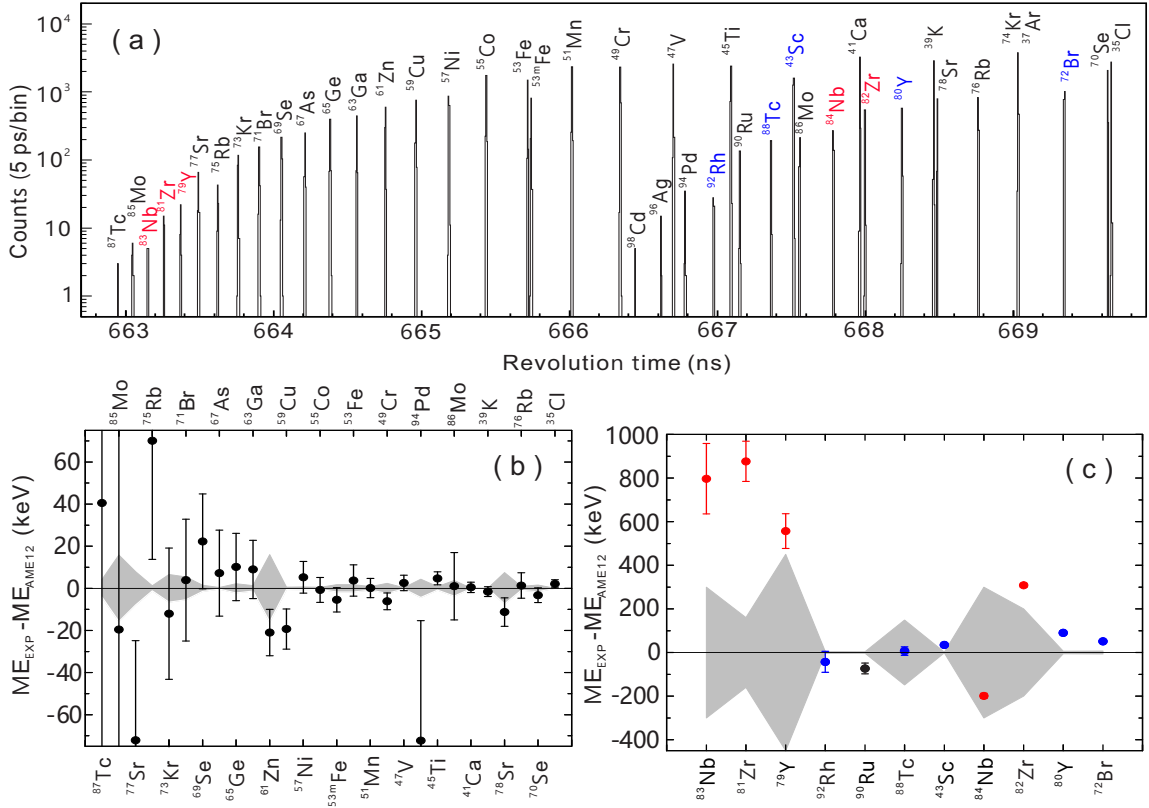}
\caption{\label{fig:T-Mass_82Zr} (Colour online)
	(a) Part of the revolution time spectrum. The nuclides of interest are indicated with red letters; the peaks with possible isomer mixture not resolved in this work are shown with blue letters.
	~Bottom: Differences between the re-determined mass excess values from this work ($ME_{exp}$) and those from AME$^{\prime}$12~\cite{Wang2012}. Note that each of the  $ME_{exp}$ values in Fig.~1(b) is re-determined by using the other 27 references, while the $ME_{exp}$ values in Fig.~1(c) are determined by using all 28 nuclides for mass calibration (see text for details). The grey shadows represent the $1\sigma$ errors from AME$^{\prime}$12.
	}
\end{figure*}

\begin{table*}
\centering
\caption{\label{tab:ME03}
 One standard deviation of the revolution times (\emph{$\sigma_T$}), counts, the mass excess (\emph{ME}) values and proton separation energies (\emph{S$_{p}$}) based on this work (IMS) and AME$^{\prime }$12~\cite{Wang2012}. 
The differences (\emph{ME}$_{\rm IMS}$-\emph{ME}$_{\rm AME^{\prime }12}$ and \emph{S$_{p}$}$_{\rm IMS}-$\emph{S$_{p}$}$_{\rm AME^{\prime }12}$) are also listed.
The symbol "${\#}$" indicates the one from at least one extrapolated values. } 
\begin{tabular}{ccccccccc}
\hline
 Atom&\emph{$\sigma_T(ps)$}&Counts &\emph{ME}$_{\rm IMS}(keV)$ &\emph{ME}$_{\rm AME^{\prime }12}(keV)$  &$\Delta\emph{ME}(keV)$&\emph{S$_{p}$}$_{\rm IMS}(keV)$ &\emph{S$_{p}$}$_{\rm AME^{\prime }12}(keV)$ &$\Delta\emph{S$_{p}$}(keV)$ \\ \hline

$^{79}$Y&2.38&34&$-57803(80)$ & $-58360(450)$ &557(457)& 1918(80)&2475(450)&$-557(457)$\\
$^{81}$Zr&2.36&26&$-57524(92)$ &$-58400(160)$&876(185)&3666(92)&4542(160)&$-876(185)$\\
$^{82}$Zr&1.12&558&$-63632(10)$&$-63940(200)^\#$&$308(200)^\#$&5209(11)&5517(200)$^\#$&$-308(200)^\#$\\
$^{83}$Nb&2.53&10&$-57613(162)$&$-58410(300)$&797(341)&1270(162)&1759(361)$^\#$&$-489(396)^\#$\\
$^{84}$Nb&1.14&407&$-61219(12)$&$-61020(300)^\#$&$-199(300)^\#$&2597(14)&2398(300)$^\#$&199(300)$^\#$\\
 \hline
$^{78}$Y&--&--&$-52397(300)^\#$&$-52530(400)^\#$&133(500)$^\#$&1883(300)$^\#$&2016(400)$^\#$&$-133(500)^\#$\\
$^{80}$Zr&--&--&$-54176(250)^\#$&$-55520(1490)$&1344(1511)$^\#$&3662(262)$^\#$&4449(1557)&$-787(1579)^\#$\\
$^{82}$Nb&--&--&$-51790(250)^\#$&$-52200(300)^\#$&410(391)$^\#$&1555(266)$^\#$&1089(340)$^\#$&466(432)$^\#$\\
$^{84}$Mo&--&--&$-53958(250)^\#$&$-54500(400)^\#$&542(472)$^\#$&3634(298)$^\#$&3379(500)$^\#$&255(582)$^\#$\\
\hline
\end{tabular}
\end{table*}

 \section{Data analysis and results}

 Fig.~\ref{fig:T-Mass_82Zr}(a) shows a part of the spectrum
 zoomed on a time window of $662.7~{\rm ns} \le t \le 669.9$~ns. From this spectrum, the average revolution time, $T$, and its standard deviation, $\sigma_T$, for each ion have been extracted.
 Most of the measured nuclides have masses known with high precision.
 We used $N_c=28$ nuclides with well-known mass values for mass calibration (Fig.~\ref{fig:T-Mass_82Zr}(b)).
A third order polynomial function of mass-to-charge ratio versus $T$ was used for the calibration.
The obtained results are listed in Table~\ref{tab:ME03}.
Since our new experimental data were included as private communications into the latest AME$^{\prime}$16~\cite{AME2016},
for comparison we use AME$^{\prime }$12~\cite{Wang2012}.

 We have re-determined the masses of each of the $N_c$ nuclides using the other $N_c-1$ ones as references.
 The differences between the re-determined mass excesses $(ME)$ and the literature ones~\cite{Wang2012} are compared in the Fig.~\ref{fig:T-Mass_82Zr}(b). The normalized $\chi_{n}$ defined as
 \begin{equation} \label{eq1}
 \chi_{n}=\sqrt{\frac{1}{n_f}\sum_{i=1}^{N_c}\frac{[(ME)_{{\rm CSRe},i}-(ME)_{{\rm AME},i}]^2}{\sigma_{{\rm CSRe},i}^2+\sigma_{{\rm AME},i}^2}}
 \end{equation}
 is calculated with $n_f=N_c$ being the number of degrees of freedom. The calculated $\chi_{n}=0.90$ is within the expected range of $\chi_{n}=1\pm 1/\sqrt{2n_f}=1\pm 0.13$ at $1\sigma$ confidence level, indicating that no additional systematic errors have to be considered.

Our measurements yield the $ME$ values of $^{82}$Zr and $^{84}$Nb for the first time, within an uncertainty of as low as $10\sim 12$ keV.
The masses of $^{81}$Zr and $^{83}$Nb are obtained to be 876(185) keV and 797(341) keV, respectively, larger than in the AME$^{\prime }$12~\cite{Wang2012}. We note that the previous $ME$ values for $^{81}$Zr as well as $^{85}$Mo are both inferred from the measurements of $\beta$-delayed proton emissions~\cite{Huang1999}.
In the case of $^{85}$Mo, a recent SHIPTRAP experiment~\cite{Haettner2011}
has shown that it is 1.59 MeV less bound than the literature value~\cite{Huang1999}.
We now show that also $^{81}$Zr is by $\sim1$ MeV less bound than the one from the same work~\cite{Huang1999}. 
Similarly, the masses of $^{83}$Nb and $^{85}$Nb were previously obtained from the $\beta$-endpoint measurements~\cite{Kuroyanagi1988}.
Both nuclei are found to be significantly less bound in a JYFLTRAP experiment  ($^{85}$Nb, by 877 keV)~\cite{Kankainen2006}
and in this work ($^{83}$Nb, by 797 keV).

\begin{figure}[!htbp]
  \centering
 	\includegraphics[width=8cm]{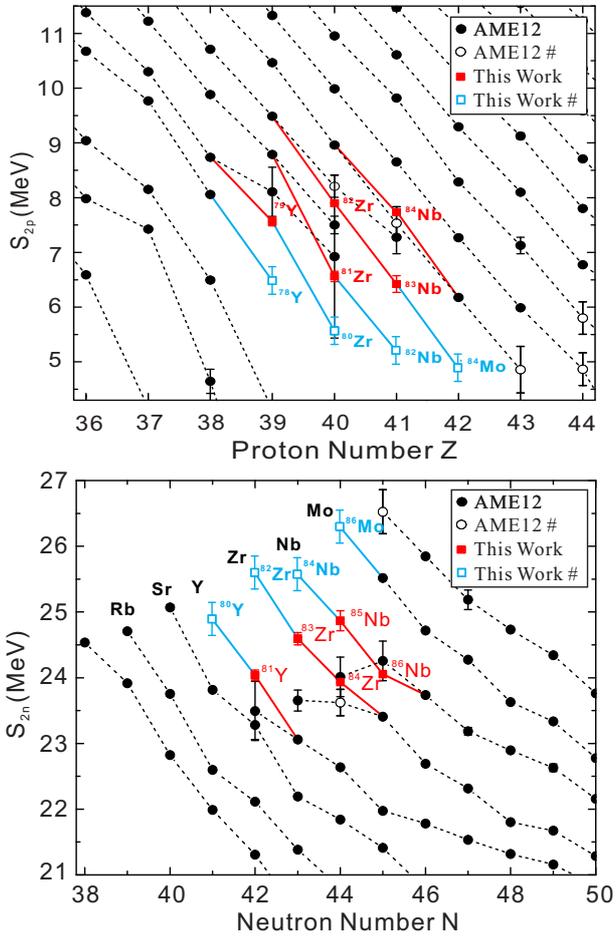}
  	\caption{\label{fig:S2p-S2n} Two-proton (\emph{S$_{2p}$}) and two-neutron (\emph{S$_{2n}$}) separation energies
	from AME$^{\prime}$12 (black) and this work (red and blue).
	Measured values are indicated by filled symbols.
	Extrapolated values (at least one from two masses is extrapolated) are indicated by open symbols. }
 \end{figure}

Fig.~\ref{fig:S2p-S2n} shows two-proton (\emph{S$_{2p}$}) and two-neutron (\emph{S$_{2n}$}) separation energies for the neutron deficient isotopes in the $A=80$ mass region. If our new mass values are used, the systematic trends of \emph{S$_{2p}$} and \emph{S$_{2n}$} become much smoother. In particular, the striking irregularities in \emph{S$_{2n}$} for $^{81}$Y, $^{83}$Zr and $^{85}$Nb (see the lower panel of Fig.~2) are removed. Using the systematics of \emph{S$_{2p}$}, \emph{S$_{2n}$}, as well as \emph{S$_{p}$} and \emph{S$_{n}$}, the masses of $^{78}$Y, $^{80}$Zr, $^{82}$Nb and $^{84}$Mo are extrapolated and averaged as given in Table~\ref{tab:ME03}. Details of this analysis will be presented in a forthcoming article.

The deviations of the re-determined $ME$s for $^{43}$Sc, $^{80}$Y, and $^{72}$Br are due to the known isomers~\cite{Wang2012} at 151-keV, 228-keV, and 100-keV excitation energies, respectively.
Isomeric states have been suggested in $^{88}$Tc~\cite{88Tc1,88Tc2} and $^{92}$Rh~\cite{92Rh1}.
Our mass value for $^{88}$Tc agrees well with the result from JYFLTRAP~\cite{Weber2008}.
We note, that the widths of the revolution time peaks of $^{88}$Tc and $^{92}$Rh follow the systematics.
This can indicate that either only one state is mainly produced in the employed nuclear reaction or the excitation energies of these isomers are very small.
We also note, that Ref.~\cite{92Rh2} did not observe the population of the isomeric state in $^{92}$Rh in fragmentation reaction.

The determined $ME$ value for $^{90}$Ru is by 73(25) keV more bound compared to the precise value in the literature~\cite{Wang2012} obtained from three independent penning trap measurements \cite{Weber2008,Fallis2011}.
The reason for this discrepancy is unknown and needs further investigation.
Using $^{90}$Ru as a calibrant does not affect the results listed in Table~\ref{tab:ME03}.

\section{rp-process}
\begin{figure}[!htbp]
\centering
\includegraphics[width=7cm]{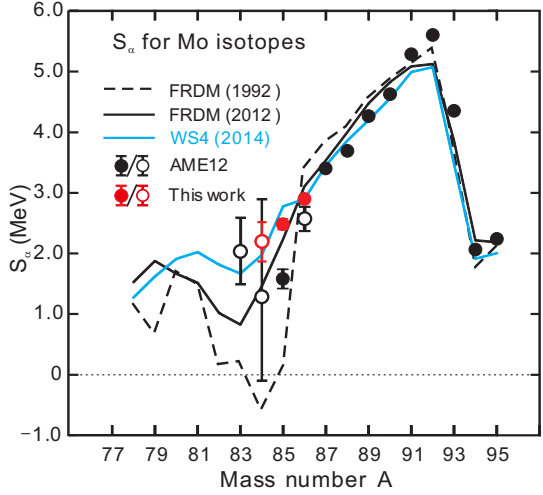}
\caption{\label{fig:mass-model} $\alpha$ separation energies, $S_{\alpha}$, for Mo isotopes. The open circle indicates $S_{\alpha}$ from at least one extrapolated mass value. The lines are from different mass models (see text).}
\end{figure}

Our new results question the pronounced island of low $\alpha $ separation energies $(S_{\alpha})$
in neutron-deficient Mo isotopes, which was predicted by FRDM$^{\prime}$92~\cite{moller95}. However such low $S_{\alpha}$ value at $^{84}$Mo was not supported by, e.g., the FRDM$^{\prime}$12~\cite{moller16} and  WS4~\cite{Wang14} mass models, though they also show a minimum in the alpha separation energies.
Figure~\ref{fig:mass-model} depicts the experimental and theoretical $S_{\alpha}$ values for Mo isotopes.
The $S_{\alpha}$ values of $^{85}$Mo and $^{86}$Mo in AME$^{\prime}$12 follow the predictions of FRDM$^{\prime}$92
if the previously known experimental mass of $^{81}$Zr and the extrapolated one of $^{82}$Zr~\cite{Huang1999,Wang2012} are used.
A sudden drop of $S_{\alpha}$ at $^{85}$Mo was called to be the first evidence of the pronounced low-$S_{\alpha}$ island~\cite{Haettner2011}.
However, if our accurate masses of $^{81,82}$Zr are used, $S_{\alpha}$ decreases smoothly with $A$ down to $^{85}$Mo
and no sudden drop of $S_{\alpha}$ at $^{85}$Mo is observed.
It is also the case for Tc isotopes, for which the reported sudden decrease of $S_{\alpha}$ at $^{87}$Tc~\cite{Haettner2011}
is now removed due to our new mass of $^{83}$Nb.
Fig.~\ref{fig:mass-model} shows that the new experimental $S_{\alpha}$ data
can be well described by the latest version of FRDM$^{\prime}$12~\cite{moller16} and WS4~\cite{Wang14} mass models.
The latter has been found to be the most accurate model in various mass regions~\cite{Sobi14,Adam18}.
We note, that the extrapolated $S_{\alpha}(^{84}{\rm Mo})$ agrees well with the prediction by the WS4 model.
The facts above indicate that the claimed pronounced low-$S_{\alpha}$ island in neutron-deficient Mo isotopes does not exist.

The non-existence of the low-$S_{\alpha}$ island in neutron-deficient Mo isotopes questions the formation of the predicted Zr-Nb cycle in the rp-process of type I X-ray bursts~\cite{Schatz98}.
Such Zr-Nb cycle is characterized by large $^{84}{\rm Mo}(\gamma,\alpha)^{80}{\rm Zr}$ and $^{83}{\rm Nb}(p,\alpha)^{80}{\rm Zr}$ reaction rates,
which sensitively depend on $S_{\alpha}(^{84}{\rm Mo})$, i.e., the mass difference between $^{84}$Mo and $^{80}$Zr.
Based on our extrapolated masses of $^{84}$Mo and $^{80}$Zr, we obtain $S_{\alpha}(^{84}{\rm Mo})=2.21(35)^{\#}$ MeV.
This value agrees with the previous extrapolations but is somewhat higher than the values used in the previous type I X-ray burst model calculations in Refs.~\cite{Schatz98,Haettner2011}.
Furthermore, it indicates that the expected large $^{84}{\rm Mo}(\gamma,\alpha)^{80}{\rm Zr}$ and $^{83}{\rm Nb}(p,\alpha)^{80}{\rm Zr}$ reaction rates
could significantly be reduced, leading to a weakening or even disappearance of the Zr-Nb cycle in the $rp$-process in type I X-ray bursts.

\begin{figure}[!htbp]
\centering
	\includegraphics[width=8cm]{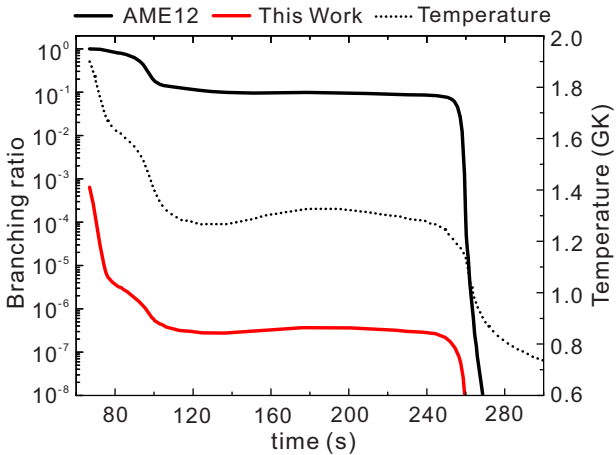}
	\caption{\label{fig:Bcycle} Fraction of the reaction flow branching into the Zr-Nb cycle
	under the most favorable conditions (see text for details) if using the masses from AME$^{\prime}$12 (black line) and from this work (red line).
	The dashed black line shows the temperature varying within the burst time. For clarity, only the cooling stage is presented.}
\end{figure}

Network calculations~\cite{Bojazi2014} based on the type I X-ray burst model of Schatz et al.~\cite{Schatz2001} have been performed using the new reaction rates obtained with the Talys code~\cite{Gori2008, Koni2008}.
We define a cycle branching ratio as  the fraction of the flow ending at $^{80}$Zr
via the $^{83}{\rm Nb}(p,\alpha)^{80}{\rm Zr}$ and $^{84}{\rm Mo}(\gamma,\alpha)^{80}{\rm Zr}$ reactions.
Calculations show that if our new results are used, the reaction rates favoring the formation of the Zr-Nb cycle 
are reduced by orders of magnitude.
Fig.~\ref{fig:Bcycle} shows the cycle branching ratio as a function of burst time for a typical burst~\cite{Schatz2001}. Under the favorable conditions we assume the $1\sigma$ upper or lower limits of mass uncertainties which give the largest $Q$-value
for the $^{83}{\rm Nb}(p,\alpha)^{80}{\rm Zr}$ reaction and the smallest $\alpha$ separation energy of $^{84}{\rm Mo}$.
If the favorable masses from AME$^{\prime}$12 are used (black solid line in Fig.~4), a large branching ratio can be found at the peak temperature of $\sim$1.9 GK, which is the same result as obtained in Ref.~\cite{Haettner2011}.
The branching ratio is reduced quickly as the temperature decreases to below 1.4 GK.
However, if our new masses are taken, the branching ratio into the Zr-Nb cycle is decreased, as demonstrated by the red line in Fig.~\ref{fig:Bcycle},
by several orders of magnitude even at the peak temperature of $\sim$1.9 GK.

Until recently it has been assumed that at high temperatures (above 2 GK) the $rp$-process flow stalls at the $^{56}$Ni waiting point. However, with the new mass measurement of $^{56}$Cu~\cite{56Cu} there might be some flow bypassing the $^{56}$Ni waiting point. Furthermore, there might be a possibility of an $rp$-process environment with seed nuclei beyond $^{56}$Ni or with slowly rising temperature to make the flow pass through $^{56}$Ni before reaching high temperatures. In such cases, the Zr-Nb cycle may have been relevant as a further hindrance until temperatures declined down to 1.7 GK. Our new results with certainty remove this barrier.




\section{$\nu$p-process}

In order to examine the effect of the new masses on the \emph{$\nu$p}-process,
we used a semi-analytic neutrino-driven wind model and the reaction network code
to obtain the thermodynamic trajectories of neutrino-driven outflows and productions of \emph{$\nu$p}-process.
More details can be found in Ref.~\cite{Wanajo2011}.
The parameters of the wind model are the "standard" ones which represent typical supernova conditions.

\begin{figure}[t]
\centering
	\includegraphics[width=8cm]{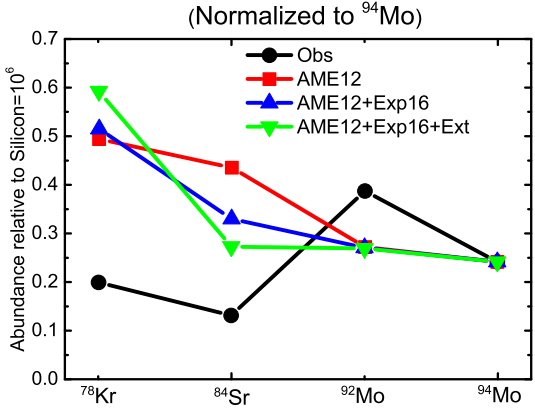}
	\caption{\label{fig:Nup_Production} Observed Solar system abundances (Obs)~\cite{Lodd2003} and \emph{$\nu$p}-process calculations based on
	 the mass values from AME$^{\prime}$12 (AME12), our new mass values (Exp16) and our extrapolated mass values (Ext).
	 Note that all the calculated abundances are normalized to $^{94}$Mo
	}
\end{figure}

Our calculations show that the new masses mainly affect the mass fractions in the mass region of $A=76\sim 86$.
In Fig.~\ref{fig:Nup_Production} we show the resulting abundances for the $p$-nuclei in this mass region.
We normalize the results to the abundance of $^{94}$Mo and compare to the solar system abundances~\cite{Lodd2003} shown as filled black circles.
The new masses affect neither $^{92}$Mo nor $^{94}$Mo.
Even though in our calculation $^{92}$Mo is a little bit more abundant than $^{94}$Mo,
it is not sufficient to explain the observed solar $^{92}$Mo/$^{94}$Mo abundance ratio,
which requires another mechanism as suggested by Wanajo~\cite{Wanajo2006}, Fisker~\cite{Fisker2009}, and Travaglio~\cite{Trav2011}.
However, the new masses have considerable effects on the production of $p$-nuclei $^{78}$Kr and $^{84}$Sr.



$^{78}$Sr is the progenitor of $^{78}$Kr. Obviously, the production of $^{78}$Sr is affected by the extrapolated masses of neighboring $^{78}$Y and $^{80}$Zr and by the measured mass of  $^{79}$Y. The relative abundance of $^{78}$Kr is slightly increased in the calculations if our new results are taken into account. Thus the overproduction of $^{78}$Kr relative to $^{94}$Mo became even stronger. This result calls for further precision mass measurements of the neighboring N=Z nuclides. Also, it gives support for the reconsideration of the significant \emph{$\nu$p}-process contribution to $^{94}$Mo abundance as suggested by Wanajo~\cite{Wanajo2011}.




Furthermore, the abundance of $^{84}$Sr, which appears overproduced with respect to the Mo isotopes in previous calculations, is reduced.
This change, which is largely related to the decrease by $\sim$500~keV of the proton separation energy of $^{83}$Nb,
modifies the reaction flow from $^{82}$Zr$(p,\gamma)^{83}$Nb$(p,\gamma)^{84}$Mo to $^{82}$Zr$(p,\gamma)^{83}$Nb$(n,p)^{83}$Zr$(p,\gamma)^{84}$Nb.
Hence, $^{84}$Nb becomes the progenitor of $^{84}$Sr if the new masses are used
instead of $^{84}$Mo if the masses from the AME$^{\prime}$12 are considered.
This change alone will lead to a substantial decrease in the production of $^{84}$Sr.
However, it is partly compensated by the increase in the proton separation energy of $^{82}$Nb.
The latter allows for the reaction sequence $^{81}$Zr$(p,\gamma)^{82}$Nb$(n,p)^{82}$Zr
feeding into the reaction chain described above.
The proton separation energy of $^{82}$Nb is based on the extrapolation.
It is rather close to the value defining the $\nu p$-process path assuming
$(p,\gamma)\rightleftarrows(\gamma,p)$ equilibrium, which is about
1.65~MeV for the conditions considered here.
An experimental mass for $^{82}$Nb would thus be highly welcome.


\section{Summary}

In summary, the masses of five neutron-deficient nuclei, $^{79}$Y, $^{81,82}$Zr, and $^{83,84}$Nb have been precisely measured using isochronous mass spectrometry at HIRFL-CSR.
Our new mass values do not support the existence of a pronounced low-$S_{\alpha}$ island in Mo isotopes.
As a consequence, the predicted Zr-Nb cycle in the $rp$-process of type I X-ray bursts does not exist or at least is much weaker than previously expected.
Furthermore, our new data allowed for elimination of some uncertainties in the \emph{$\nu$p}-process induced by the poorly-known nuclear masses.
Based on our new mass values, the abundance estimation of the \emph{$\nu$p}-process to the $p$-nuclides in the A$\sim$90 region
lies now on a more solid basis in terms of masses, although there are other yet unknown physical parameters, such as $(n,p)$ reaction rates.
Particularly, the new masses lead to a reduction of the $^{84}$Sr abundance.
This reduces the overproduction of $^{84}$Sr relative to $^{92,94}$Mo which was found
in the previous \emph{$\nu$p}-process calculations~\cite{Wanajo2011,Frohlich2006,Pruet2006}.
Our study also implies that additional important mechanisms beyond the \emph{$\nu$p}-process, such as the effect of neutrino oscillations~\cite{PRD2,PRD1,PRD}, are needed to explain the observed $p$-nuclei abundances.

\section*{Acknowledgments}
We thank also the staffs in the accelerator division of IMP for providing stable beam.
This work is supported in part by the National Key Program for S\&T Research and Development (Contract No. 2016YFA0400504), the Major State Basic Research Development Program of China (Contract No. 2013CB834401), the Key Research Program of Frontier Sciences, CAS (Grant No. QYZDJ-SSW-S),
the National Natural Science Foundation of China (NSFC)
(Grants No. U1232208, U1432125, 11205205, 11035007, 11235001, 11320101004, 11575007, 11605252, 11711540016, and 11575112),
the Helmholtz-CAS Joint Research Group (Grant No. HCJRG-108), and
the European Research Council (ERC) under the European Union's Horizon 2020 research and innovation programme (grant agreement No 682841 ``ASTRUm'').
S.W. acknowledges support by the RIKEN iTHES Project and the JSPS Grants-in-Aid for Scientific Research (Grant No. 26400232 and 26400237).
Y.A.L. acknowledges support by the CAS President's International Fellowship Initiative Grant (Grant No. 2016VMA043).
Y.H.L. thanks for support from the Ministry of Science and Technology of China (Talent Young Scientist Program) and from the China Postdoctoral Science Foundation (Grant No. 2014M562481).
Y.H.Z. acknowledges support by the ExtreMe Matter Institute EMMI at the GSI Helmholtzzentrum f{\"u}r Schwerionenforschung, Darmstadt, Germany.
X.X. thanks for support from CAS "Light of West China" Program.
A.S. and G.M.P. acknowledge support from the Deutsche Forschungsgemeinschaft through contract SFB 1245.
K.B. and G.M.P. thank for support from the Nuclear Astrophysics Virtual Institute (NAVI) of the Helmholtz Association.
C.F. acknowledges support by the U.S. Department of Energy, Office of Science under Award Number SC0010263.



\begin{thebibliography} {99}
    \bibitem{Rauscher2014} T.~Rauscher, Aip. Adv. {\bf 4}, 041012 (2014).
    \bibitem{Kajino2016} C.~A.~Bertulani and T.~Kajino, Prog.~Part.~Nucl.~Phys. {\bf 89}, 56 (2016).
	\bibitem{Kappeler2011} F.~Kappeler, R.~Gallino, S.~Bisterzo, and W.~Aoki, Rev.~Mod.~Phys.~{\bf 83}, 157 (2011).
	\bibitem{Arnould2007} M.~Arnould, S.~Goriely, and K.~Takahashi, Phys.~Rep. {\bf 450}, 97 (2007).
	\bibitem{Arnould2003} M.~Arnould and S.~Goriely, Phys.~Rep.~{\bf 384}, 1 (2003).
	\bibitem{Rauscher2002} T.~Rauscher {\it et al.}, Astrophys.~J.~{\bf 576}, 323 (2002).
	\bibitem{Wallace1981} R.~K.~Wallace and S.~E.~Woosley, Astrophys.~J.~Suppl.~Ser.
	{\bf 45}, 389 (1981).
	\bibitem{Frohlich2006} C.~Fr\"{o}hlich {\it et al.}, Phys.~Rev.~Lett.~{\bf 96}, 142502 (2006).
	\bibitem{Wanajo2006} S.~Wanajo, Astrophys.~J.~{\bf 647}, 1323 (2006).
	\bibitem{Pruet2006} J.~Pruet {\it et al.}, Astrophys.~J.~{\bf 644}, 1028 (2006).
	\bibitem{Trav2011}C.~Travaglio et al., ApJ {\bf 739}, 93 (2011)
	\bibitem{Woosley1990} S.~E.~Woosley {\it et al.}, Astrophys.~J.~{\bf 356}, 272 (1990).
    \bibitem{Wanajo2011} S.~Wanajo, H.-Th. Janka and S. Kubono, Astrophys.~J.~{\bf 729}, 46 (2011).  	
	\bibitem{Schatz98} H.~Schatz {\it et al.}, Phys.~Rep.~{\bf 294}, 167 (1998).
	\bibitem{Schatz2001} H.~Schatz {\it et al.}, Phys.~Rev.~Lett.~{\bf 86}, 3471 (2001).
	\bibitem{Elomaa} V.-V.~Elomaa {\it et al.}, Phys.~Rev.~Lett.~{\bf 102}, 252501 (2009).
	\bibitem{Parikh2013} A.~Parikh {\it et al.}, Prog.~Part.~Nucl.~Phys. {\bf 69}, 225 (2013).

	\bibitem{Arcones1} A.~Arcones, C. Fr{\"o}hlich and G. Mart{\'i}nez-Pinedo, Astrophys.~J.~{\bf 750}, 18 (2012).

    \bibitem{Parikh2009} A.~Parikh {\it et al.}, Phys.~Rev.~C {\bf 79}, 045802 (2009).	
	\bibitem{Gabriel-Book} G. Mart{\'i}nez-Pinedo {\it et al.} in Handbook of Supernovae (Springer International Publishing,
Cham, 2016), pp. 1--37.
	\bibitem{Schatz2016} H.~Schatz and W.-J. Ong, Astrophys.~J.~{\bf 844}, 139 (2017).

	\bibitem{moller95} P.~M\"{o}ller {\it et al.}, At.~Data~Nucl.~Data~Tables {\bf 59}, 185 (1995).
	\bibitem{Haettner2011} E.~Haettner {\it et al.}, Phys.~Rev.~Lett.~{\bf 106}, 122501 (2011).
	
	\bibitem{Fallis2008} J.~Fallis {\it et al.}, Phys.~Rev.~C {\bf 78}, 022801 (2008).
	\bibitem{Weber2008} C.~Weber {\it et al.}, Phys.~Rev.~C {\bf 78}, 054310 (2008).
	\bibitem{Fisker2009} J.~L.~Fisker {\it et al.},
	Astrophys.~J.~Lett.~{\bf 690}, L135 (2009).
	\bibitem{Fallis2011} J.~Fallis {\it et al.}, Phys.~Rev.~C {\bf 84}, 045807 (2011).
	
	
	
	\bibitem{Audi2003} G.~Audi, A.~H.~Wapstra, and C.~Thibault, Nucl.~Phys.~{\bf A729}, 337 (2003).

	\bibitem{Xiajiawen} J.~W.~Xia {\it et al.},
	Nucl.~Instrum.~Meth.~A {\bf 488}, 11 (2002).
	\bibitem{zwl2010} W.~L.~Zhan {\it et al.},	
	Nucl.~Phys.~A~{\bf 834}, 694c (2010).
	
	\bibitem{Tu2011}X.~L.~Tu {\it et al.}, Nucl.~Instrum.~Meth.~A~{\bf 654}, 213 (2011).
	\bibitem{Shuai-MFC} P. Shuai \emph{et al.}, arXiv:1407.3459
	\bibitem{Chen2017} R. J. Chen {\it et al.}, Comp. Phys. Comm. {\bf 221}, 216 (2017).
	
	\bibitem{Geis92} H.~Geissel {\it et al.},
	Nucl.~Instrum.~Meth.~B {\bf70}, 286 (1992).



\bibitem{ims1} M. Hausmann {\it et al.}, Nucl. Instrum. Meth. A {\bf 446}, 569 (2000).
\bibitem{ims2} B. Franzke {\it et al.}, Mass Spectrom. Rev. {\bf 27}, 428 (2008).
\bibitem{ims3} F. Bosch {\it et al.}, Prog. Part. Nucl. Phys. {\bf 73}, 84 (2013).
\bibitem{ims4} Y. H. Zhang {\it et al.}, Phys. Scripta {\bf 91}, 073002 (2016).

	\bibitem{Mei2010}B.~Mei {\it et al.}, Nucl.~Instrum.~Meth.~A~{\bf 624}, 109 (2010).


	
	\bibitem{AME2016}
	G.~Audi {\it et al.}, Chin. Phys. C \textbf{41}, 030001 (2017);\\
	M.~Wang {\it et al.}, Chin. Phys. C \textbf{41}, 030003 (2017).
	\bibitem{Wang2012} M.~Wang {\it et al.},
	Chin. Phys. C {\bf 36}, 1603 (2012);\\
	G.~Audi {\it et al.},
	Chin. Phys. C {\bf 36}, 1157 (2012).	
	
	\bibitem{Huang1999}W.~X~Huang {\it et al.}, Phys.~Rev.~C~{\bf 59}, 2402 (1999).
	\bibitem{Kuroyanagi1988}T.~Kuroyanagi {\it et al.}, Nucl.~Phys.~A~{\bf 484}, 264 (1988).
	\bibitem{Kankainen2006}A.~Kankainen {\it et al.}, Eur.~Phys.~J.~A~{\bf 29}, 271 (2006).
	\bibitem{88Tc1} A. Odahara {\it et al.}, Z. Phys. A {\bf 354}, 231 (1996).
	\bibitem{88Tc2} A.B. Garnsworthy {\it et al.}, Phys. Rev. C {\bf 80}, 064303 (2009).
	\bibitem{92Rh1} S. Dean {\it et al.}, Eur. Phys. J. A {\bf 21}, 243 (2004).
	\bibitem{92Rh2} G. Lorusso {\it et al.}, Phys. Rev. C {\bf 86}, 014313 (2012).

	\bibitem{moller16} P.~M\"{o}ller {\it et al.}, At.~Data~Nucl.~Data~Tables {\bf 109-110}, 1 (2016).
	\bibitem{Wang14} N.~Wang, M.~Liu, X.~Z.~Wu, and J.~Meng, Phys.~Lett.~B~{\bf 734}, 215 (2014).
	\bibitem{Sobi14} A. Sobiczewski and Yu. A. Litvinov, Phys.~Rev.~C {\bf 89}, 024311 (2014).
	\bibitem{Adam18} A. Sobiczewski {\it et al.}, At. Data Nucl. Data Tables {\bf 119}, 1 (2018).
	\bibitem{Bojazi2014} M. J. Bojazi and B. S. Meyer, Phys.~Rev.~C {\bf 89}, 025807 (2014).
	\bibitem{Gori2008} S.~Goriely, S.~Hilaire, and A.~J.~Koning, Astron and Astrophys {\bf 487}, 767 (2008).
	\bibitem {Koni2008} A.~J.~Koning, S.~Hilaire, and M.~C.~Duijvestijn, International Conference on Nuclear Data for Science and Technology, Vol 1, Proceedings, 211 (2008).
\bibitem {Lodd2003} K.~Lodders, Astrophys J {\bf591}, 1220 (2003).
	\bibitem{56Cu} A.~A.~Valverde {\it et al.}, Phys. Rev. Lett. {\bf 120}, 032701 (2018).
	\bibitem{PRD2} R. D. Hoffmann {\it et al.}, Astrophys. J. {\bf 460}, 478 (1996).
	\bibitem{PRD1} M. R. Wu {\it et al.}, Phys. Rev. D {\bf 91}, 065016 (2015).
	\bibitem{PRD} H. Sasaki {\it et al.}, Phys. Rev. D {\bf 96}, 043013 (2017).
	
		
	
\end{thebibliography}
\end{document}